%% file: papers.tex
\documentclass[runningheads]{llncs}
\usepackage{amsmath}
\usepackage{amsfonts}

\usepackage{graphicx}
\usepackage{url}
\usepackage[subrefformat=parens,labelformat=parens]{subfig}
\usepackage{enumitem}
\setlist{ 
	itemsep=0mm,
	parsep=0mm,
	topsep=1mm,
}

\usepackage[T1]{fontenc}
\usepackage[scaled=0.85]{beramono}
\usepackage{listings}
\usepackage{ctable}
\usepackage{multirow}

\newcommand{\ttl}{SPARQL with XQuery-based Filtering}
\newcommand{\atr}{Takahiro Komamizu}

\usepackage{rotating}

\begin{document}

\title{\ttl\thanks{This is a full version of the poster paper in ISWC 2020.}}
%\titlerunning{Abbreviated paper title}

\author{\atr}
%\institute{Anonymous}
\authorrunning{T. Komamizu}
\institute{Nagoya University, Japan\\
\email{taka-coma@acm.org}}

\maketitle              % typeset the header of the contribution
\begin{abstract}
\input{src/abst.tex}
%\keywords{SPARQL Extension \and XQuery Filtering \and Query Processing} 
\end{abstract}

\renewcommand{\thelstlisting}{\arabic{lstlisting}}

\input{src/intro.tex}

\input{src/rel.tex}
\input{src/def.tex}
\input{src/prop.tex}

\input{src/exp.tex}

\input{src/con.tex}

\bibliographystyle{splncs04}
\bibliography{papers}

\end{document}

%% file: src/abst.tex
Linked Open Data (LOD) has been proliferated over various domains,
however, there are still lots of open data in various format
other than RDF, a standard data description framework in LOD. 
These open data can also be connected to entities in LOD when 
they are associated with URIs.
Document-centric XML data are such open data that are connected 
with entities in LOD as supplemental documents for these entities,
and to convert these XML data into RDF requires various techniques 
such as information extraction, ontology design and ontology mapping
with human prior knowledge.
To utilize document-centric XML data linked from entities in LOD,
in this paper, a SPARQL-based seamless access method on RDF and 
XML data is proposed.
In particular, an extension to SPARQL, \texttt{XQueryFILTER}, 
which enables XQuery as a filter in SPARQL is proposed.
For efficient query processing of the combination of SPARQL and 
XQuery, a database theory-based query optimization is proposed.
Real-world scenario-based experiments in this paper showcase 
that effectiveness of \texttt{XQueryFILTER} and efficiency of 
the optimization.

%% file: src/intro.tex
\section{Introduction}

Open data movement is a worldwide movement that data are published 
online with FAIR principle\footnote{FAIR stands for findability, accessibility,
interoperability and reusability}.
Linked Open Data (LOD)~\cite{linked_data} started by Sir Tim Berners-Lee
is best aligned with this principle.
In LOD, factual records are represented by a set of triples consisting 
of subject, predicate and object in the form of a standardized representation 
framework, RDF (Resource Description Framework)~\cite{rdf}.
Each element in RDF is represented by network-accessible identifier 
called URI (Uniform Resource Identifier).
SPARQL~\cite{sparql} is a standard query language for RDF data, and
its basic structure is a graph pattern which consists of triple patterns
being matched with triples in RDF data.

Although LOD has been proliferated its population over various domains,
there are still lots of data which require large effort on converting into LOD.
In particular, documents such as manuals and minute books are in this line. 
To convert document structure into RDF is one choice, however, the converted RDF 
schema and its data are not semantically meaningful.
To utilize contents of documents, extracting facts from documents
in the form of triples is another choice.
To this end, sophisticated techniques (information extraction, ontology design,
ontology mapping, and more) have been studied, but they still require human
efforts on error-pruning.

In this paper, the way to utilize document open data which are linked from
entities in LOD is explored.
A na\"{i}ve approach is a step-by-step approach.
It first process SPARQL instance to obtain entities with links to documents.
Second, these documents are processed on the basis of users' purpose such as 
filtering and analysis. 
In most cases, full-length documents are not necessary 
but only parts of documents are required.
If these documents are structured (e.g., XML or JSON) and stored in
dedicated databases (e.g., XML DB or document DB), their queryable interface
allows users to extract parts of documents.

Consider the following example scenario.
A user attempts to survey the proliferation of country-wise warnings about a virus
with respect to county populations.
Suppose that national safety information are published in XML and county information 
are available on LOD such as DBpedia.
She first searches on the national safety information by using XQuery for countries that 
warnings about the virus are announced in a specified period.
For each country in the XQuery results, she searches DBpedia for
finding its population by using SPARQL.
At last, she analyses these results for finding relationships
between population and warning proliferation.
In summary, this user is required three processes, namely, XQuery processing, SPARQL 
processing and merging results of different formats (i.e., XML and tuple).
This example indicates that utilizing entities in LOD with documents is laborious
that communication with different interfaces and
merging results of various formats are required.

In this paper, to realize efficient query processing over LOD with documents,
a SPARQL-based seamless access method is proposed.
In particular, XML is assumed as the format of documents, 
therefore, the proposed method combines XQuery into SPARQL.
For the first attempt, the usage of XQuery in SPARQL is limited for filtering.
To this end, a dedicated filtering expression, \texttt{XQueryFILTER}, 
for SPARQL is proposed in this paper.
This expression enables XQuery-based filtering by accessing underlying XML data.
There are basically three query execution plans with \texttt{XQueryFILTER} as follows.
\textbf{Parallel plan} is a plan that SPARQL and XQuery are performed in parallel and 
their results are joined afterward.
\textbf{SPARQL first plan} is a plan that SPARQL is performed first, bindings of SPARQL variables
are pushed into XQuery, and XQuery is performed.
\textbf{XQuery first plan} is a plan that XQuery is first performed with modification to
return document identifiers and SPARQL with these identifiers is performed.
To explore the optimal plans, in this paper, database theory-based query optimization~\cite{database} 
which utilizes statistics of selectivity of SPARQL and XQuery as well as 
performance of SPARQL endpoint and XML DB is proposed.

The contributions of this paper are summarized as follows.
\begin{itemize}
	\item \textbf{Novel Filtering on LOD}:
		In this paper, to utilize documents connected from entities in LOD,
		SPARQL is extended to enable document-based filtering.
		In particular, the proposed method assumes XML documents as the underlying documents
		which are stored in XML DB in which XQuery can be performed.
		The proposed extension, \texttt{XQueryFILTER}, is a filtering expression in SPARQL
		that contains an XQuery instance as filtering condition.
		This XQuery instance contains a SPARQL variable which bindings are links to XML documents in XML DB.
	\item \textbf{Query Optimization}:
		SPARQL with \texttt{XQueryFILTER} contains both SPARQL and XQuery instances 
		and they can be performed separately, therefore, three possible plans can be considered,
		namely, parallel plan, SPARQL first plan and XQuery first plan.
		On the basis of query optimization in database theory,
		the cost models of these plans are designed and the optimal query plan is explored
		by using selectivity estimation and database performance statistics.
	\item \textbf{Real-world Scenario-based Experiment}:
		To demonstrate the usability of \texttt{XQueryFILTER}, three real-world scenarios
		are examined in experimental evaluation.
		The first one is country search from DBpedia using safety information XML data in 
		external database.
		The second one is law search based on law body that laws are entities in 
		law history LOD~\cite{ckg2019} and law bodies are XML documents.
		The third one is discussion search with minute book that discussions are 
		associated with laws in the law history LOD and minute books are XML documents.
		On these scenarios, the experiment showcases that the query optimization 
		can discover the optimal plans.
\end{itemize}

The rest of this paper is organized as follows.
First, related work to the proposed method is introduced in Section~\ref{sec:rel}.
In Section~\ref{sec:def}, definition of \texttt{XQueryFILTER} and 
basic architecture for \texttt{XQueryFILTER} are explained.
The proposed method for realizing \texttt{XQueryFILTER} including
query plans and optimization are discussed in Section~\ref{sec:prop}.
In Section~\ref{sec:exp}, real-world scenario-based experiments are introduced
to showcase characteristics of these plans and effects of the optimization.
Finally, conclusion and future directions are described in Section~\ref{sec:con}.

%% file: src/rel.tex
\section{Related Work}
\label{sec:rel}

In this paper, SPARQL with XQuery-based filtering is proposed, and,
there are few work on combining queries for XML (like XPath, XSLT and XQuery) into SPARQL query
except \cite{xpath_in_sparql}.
Droop et al.~\cite{xpath_in_sparql} have proposed a translation-based XPath embedding 
for SPARQL in order to enable XPath processing in SPARQL query processor.
To this end, they have proposed translation model from XML to RDF and from XPath to SPARQL.
There are three major differences from their work to the proposed filtering in this paper.
One is query type for XML, namely, XQuery and XPath.
In general, XQuery is more expressive than XPath.
Another difference is that, in the proposed approach, no preprocessing is applied to XML data, 
while \cite{xpath_in_sparql} requires translation into RDF.
The other difference is that, the proposed approach fully utilizes the dedicated 
query processing technique in XML DB, but \cite{xpath_in_sparql} translates an XPath 
instance into a complicated SPARQL query and processes it on a SPARQL query processor.

The related context is to query XML and RDF which can be translated each other.
So-called data-centric XML data~\cite{document-centric-xml} are designed for representing
objects with hierarchical attributes, while document-centric (a.k.a. content-oriented)
XML data preserve document structures where the data are still understandable without XML tags.
Therefore, data-centric XML data are easy to convert to RDF,
while document-centric XML data require large efforts on designing ontologies and mappings.
The existing approaches basically assume data-centric XML data and RDF data as their translation.
These approaches can be roughly classified into the following three categories.
(1) to use XQuery to query on RDF data, 
(2) to use SPARQL to query on XML data,
and (3) query translation between XQuery and SPARQL.

In the first category, there are two basic approaches. 
One is to integrate SPARQL into XQuery and the other is to convert RDF into XML.
One early work~\cite{xquery4lod_0} is the former approach, which aims at reshaping
the output of SPARQL queries into XML for subsequent applications.
It embeds SPARQL query into XQuery/XSLT for reconstructing SPARQL query results into XML.
A more recent work~\cite{xquery4lod_1} is the latter approach, which
aims to integrate Open Street Map (OSM) in XML and Linked Geo Open Data (LGOD).
To this end, this work has proposed XOSM (XQuery for OSM) framework
which enables XQuery over LGOD and OSM by transforming LGOD data into OSM format.
Therefore, this work does not handle RDF data directly.

In the second category,
XSPARQL~\cite{xsparql} is a representative approach, which is an extension of XQuery
for enabling querying to either XML data or RDF data by switching query expressions.
The motivation of this work is that translating XML data into RDF data is a tedious and 
error-prone task.
To deal with this situation, XSPARQL enables to query XML data to produce RDF data 
(\textit{lifting}) and to query RDF data to produce XML data (\textit{lowering}).
Note that, in XSPARQL, users can perform query on either XML data or RDF data, meaning that 
they cannot perform query both of them at the same time.

In the last category,
translating from one query language to another has been studied~\cite{translation_survey}
to enable universal access to data in any format.
In the users' perspective, learning different query languages is troublesome. 
Query translation aims to reduce this learning cost.
SPARQL2XQuery~\cite{sparql2xquery} is the only work on the last category
for translating SPARQL to XQuery.
It transforms XML schema into OWL ontology to generate mapping 
between XML data and this ontology, and on the basis of this ontology,
SPARQL query is translated into XQuery.

%% file: src/def.tex
\section{SPARQL with XQuery-based Filtering}
\label{sec:def}

In this section, the proposed extension of SPARQL is defined.
In the first sections, XQuery and SPARQL with examples are briefly overviewed.
In the following sections the definition of the proposed extension and 
a basic architecture of mixed databases (SPARQL endpoint and XML database) are introduced.

\subsection{XQuery: Brief Review}

The following gives a brief review of XQuery, and
the full definition of XQuery is referred to \cite{xquery}.
The core of XQuery is \texttt{FLWR} expression, each of which character refers 
to keywords \texttt{FOR}, \texttt{LET}, \texttt{WHERE} and \texttt{RETURN}, respectively.
\texttt{FOR} and \texttt{LET} clauses are used for binding XML elements 
to variables, \texttt{WHERE} clause serves a filter for the bindings,
and \texttt{RETURN} clause reshapes the bindings in arbitrary XML format.

Listing~\ref{lst:xquery} is an XQuery example.
The target XML data are obtained from Overseas Travel Safety Information Open Data
by Ministry of Foreign Affairs (MOFA) of 
Japan\footnote{\url{https://www.ezairyu.mofa.go.jp/html/opendata/index.html} (in Japanese)}.
The data contain safety information (e.g., risk levels and announcements) for each foreign countries.
The \textit{mail} tag in the XML data correspond to announcements from MOFA.
The XQuery in Listing~\ref{lst:xquery} is a query to search for mail tags
about a country (each XML file corresponds to a country) produced after 
March, 2020 talking about COVID-19 (i.e., coronavirus).

%\begin{figure}[t]
\begin{lstlisting}[
	mathescape=true,
	caption=XQuery Example on Travel Safety Information Open Data,
	label=lst:xquery
]
 FOR $\$$d in doc('safety_info.xml')//mail
 WHERE $\$$d/leaveDate > xs:date('2020-03-01') and contains($\$$d, 'coronavirus')
 RETURN $\$$d
\end{lstlisting}
%\end{figure}

\subsection{SPARQL: Brief Review}

The following gives a brief review of SPARQL, and
the full definition of SPARQL should be referred to \cite{sparql}.
The core of SPARQL is \textit{basic graph pattern} which consists of 
a set of \textit{triple patterns}.
A triple pattern is represented by triplet $(s, p, o)$.
$s$ and $o$ can be variable, URI or literal,
and $p$ can be either variable or URI.
Main keywords in SPARQL are \texttt{SELECT} and \texttt{WHERE}.
\texttt{WHERE} clause includes basic graph pattern and filtering conditions.
\texttt{SELECT} returns bindings of specified list of variables in \texttt{WHERE} clause.
\texttt{SERVICE} clause is an extension in SPARQL 1.1~\cite{sparql11}, which supports 
queries that merge data distributed across the Web.

Listing~\ref{lst:sparql} is a SPARQL example using \texttt{SERVICE} clause on
DBpedia\footnote{\url{http://dbpedia.org/}}.
The local SPARQL endpoint includes a mapping table between countries in
the safety information by MOFA and those in DBpedia.
The SPARQL instance in Listing~\ref{lst:sparql} is a query to search for countries 
in the local SPARQL endpoint which population is more than 10 million using 
external knowledge about countries in DBpedia.
Country identifiers in the local endpoint are explored in the \texttt{WHERE} clause, 
and external information in DBpedia to explore countries with more than 10 million
population is used in the query by using the \texttt{SERVICE} clause.

\begin{figure}[t]
\begin{lstlisting}[
	mathescape=true,
	caption=SPARQL Example on DBpedia,
	label=lst:sparql
]
  SELECT ?s 
  WHERE { ?s rdf:type ex:Country ; owl:sameAs ?x .
     SERVICE <http://dbpedia.org/sparql> {
        SELECT ?x
        WHERE { ?x dbo:populationTotal ?pop .
                FILTER ( ?pop > 10,000,000 ) .
     } }
  }
\end{lstlisting}
\end{figure}

\subsection{Query Definition}

XQuery-based filtering aims to filter bindings of basic graph pattern in SPARQL
that XML documents specified in the bindings satisfy conditions in XQuery.
To this end, this XQuery instance is required to return boolean value.
In the extended query, \texttt{XQueryFILTER} expression is added to 
the SPARQL definition, that imitates \texttt{FILTER} expressions.
Argument of \texttt{XQueryFILTER} is XQuery expression including a single variable 
used in basic graph pattern in SPARQL.
Note that SPARQL variable in XQuery is easily distinguishable from 
XQuery variable by the prefix symbol (i.e., \texttt{?} for SPARQL variable and 
\texttt{\$} for XQuery variable).

Listing~\ref{lst:xfilter} is an example of SPARQL with XQuery-based filtering.
This query is to search for countries with more that 10 million people 
and with safety information about the coronavirus after March, 2020. 
\texttt{XQueryFILTER} in the query contains a SPARQL variable \texttt{?doc} which  
referred by \texttt{ex:safetyInfo} predicate which range is URI of safety information XML file.
During SPARQL evaluation, bindings for \texttt{?doc} are determined based on the basic graph pattern.
The XQuery in the \texttt{XQueryFILTER} can be performed by replacing \texttt{?doc} variable
with one of its bindings.
If the XQuery returns \texttt{true}, the binding remains in results, 
otherwise, the binding is eliminated from results.  

%\begin{figure}[t]
\begin{lstlisting}[
	mathescape=true,
	caption=Example of SPARQL with XQuery-based Filtering,
	label=lst:xfilter
]
  SELECT ?s
  WHERE { ?s rdf:type ex:Country ; ex:safetyInfo ?doc ; owl:sameAs ?x .
    SERVICE <http://dbpedia.org/sparql> {
       SELECT ?x
       WHERE { ?x dbo:populationTotal ?pop .
               FILTER ( ?pop > 10,000,000 ) . 
    } }
    XQueryFILTER (
       LET $\$$x := doc(?doc)//mail[leaveDate > xs:date('2020-03-01')]
       RETURN contains($\$$x, 'coronavirus')
    ) . 
  }
\end{lstlisting}
%\end{figure}

\subsection{System Architecture}

\begin{figure}[t]
	\centering
	\includegraphics[width=.6\textwidth]{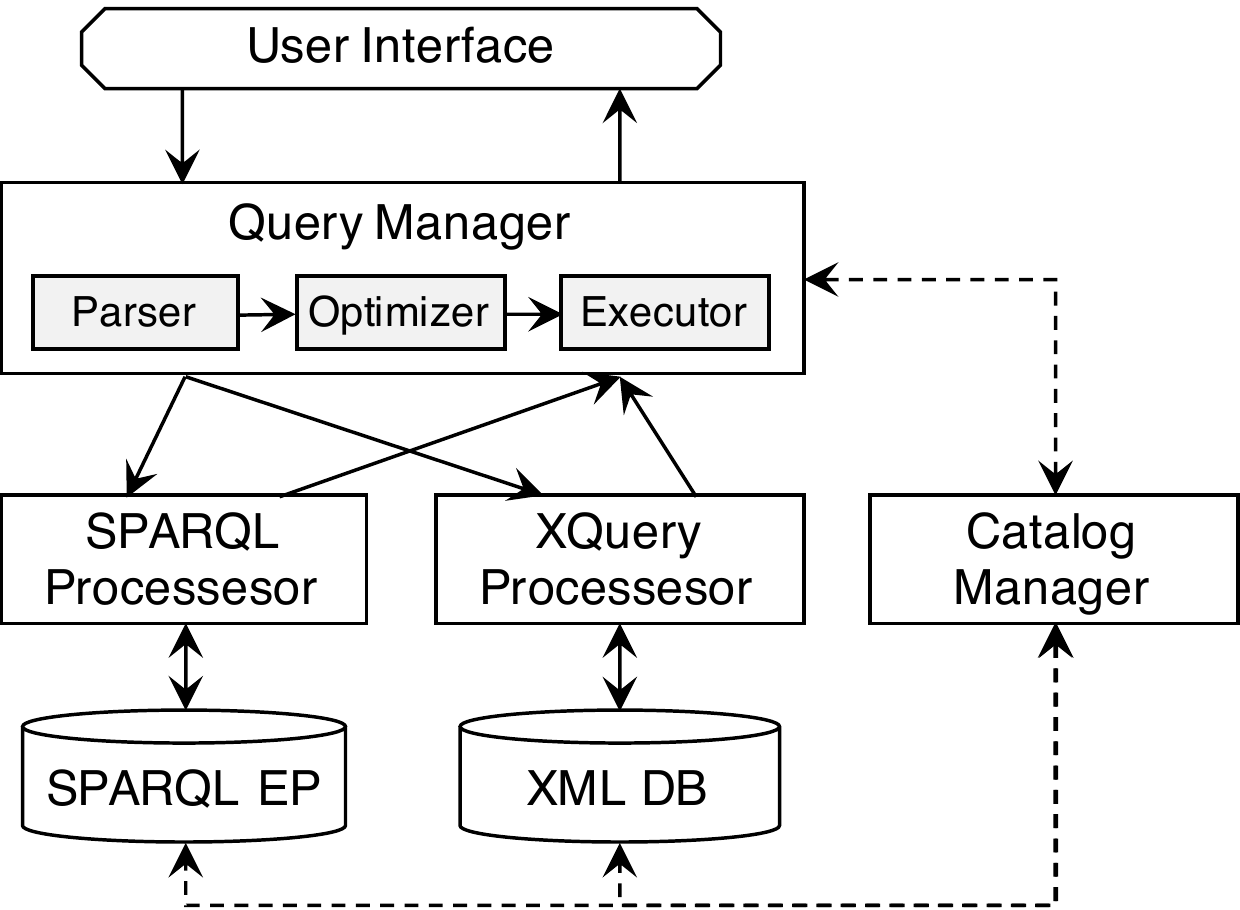}
	\caption{Basic Architecture}
	\label{fig:architecture}
\end{figure}

Figure~\ref{fig:architecture} represents a basic architecture for realizing
SPARQL with XQuery-based filtering.
A basic assumption is that RDF and XML data are separately stored in SPARQL EP (endpoint)
and XML DB, respectively.
Dedicated query processors are associated to communicate with the databases,
namely, SPARQL processor and XQuery processor.
Query manager handles user query, decomposes it into SPARQL and XQuery instances,
explores optimal query plans, executes these instances and merges the results.
User interface communicates with users by receiving queries and returns their results.

In this architecture, the query manager is responsible for query optimization.
In the parser module, it first parses input query into SPARQL and XQuery instances.
In the optimizer module, it explores optimal query execution plan of the basis of 
statistics and selectivity estimation by using the catalog manager.
The catalog manager is responsible to mange supplemental information for query optimization
by communicating to the databases.
At last, the executor module executes the decomposed queries and reshapes their results.

%% file: src/prop.tex
\section{Query Processing}
\label{sec:prop}

To realize efficient SPARQL with XQuery-based filtering,
the query manger in Figure~\ref{fig:architecture} is a core component of query processing.
This component has three subcomponents, namely, parser, optimizer and executor.
Since RDF data and XML data are separately managed,
the role of the parser is to decompose a user query into a SPARQL instance and an XQuery instance,
and, besides, to extract SPARQL variable in \texttt{XQueryFILTER} for joining results of these queries.

\begin{figure}[t]
	\centering
	\includegraphics[width=.75\textwidth]{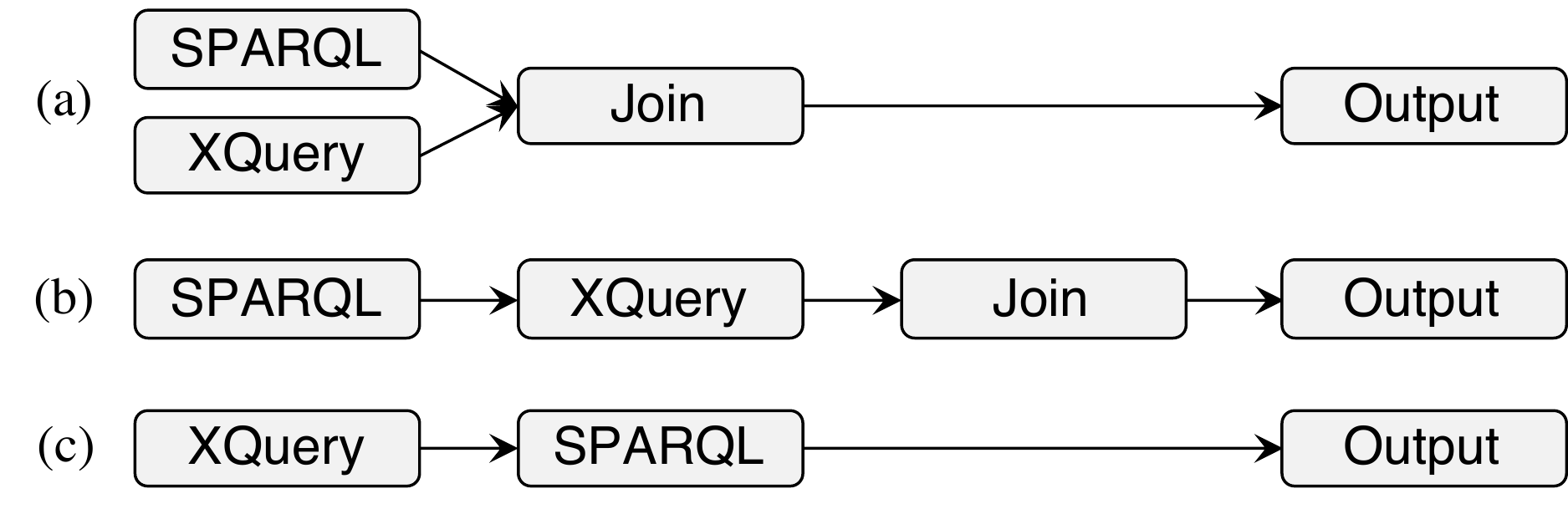}
	\caption{
		Query Plans;
		(a) Parallel,
		(b) SPARQL First, and 
		(c) XQuery First
	}
	\label{fig:manager}
\end{figure}

In the optimizer, there are three possible plans to execute SPARQL with XQuery-based filtering as 
shown in Figure~\ref{fig:manager}.
\begin{enumerate}
	\item \textbf{Parallel}: Execute SPARQL and XQuery in parallel
		and join these results afterward 
		(Figure~\ref{fig:manager} (a)).
	\item \textbf{SPARQL First}: Execute SPARQL first, push its bindings down
		into XQuery, and evaluate the bindings with XQuery results
		(Figure~\ref{fig:manager} (b)).
	\item \textbf{XQuery First}: Execute XQuery first and push its results down into SPARQL
		(Figure~\ref{fig:manager} (c)).
\end{enumerate}
Subsequent sections introduce how to realize these query execution 
plans followed by a section describing optimal query plan exploration.

\subsection{Parallel}

Basic idea of this query plan is that SPARQL endpoint and XML DB are independent 
systems, therefore, executing SPARQL and XQuery in parallel can fully utilize their capabilities.
In this plan, SPARQL and XQuery instances after decomposition requires 
query rewriting in order to execute correctly and join process afterward.
In SPARQL instance, To join results of SPARQL and XQuery, variables for join keys are 
required, thus, in SPARQL instance, the SPARQL variable appearing in XQuery instance is 
ensured to exist in \texttt{SELECT} clause.
In XQuery instance, a SPARQL variable within it leads syntax error.
Therefore, for XQuery execution, XQuery instance is rewritten to 
another instance which finds XML documents satisfying the all conditions in the queries.

In this paper, rule-based rewriting is applied to rewrite an XQuery instance. 
First, enumeration of XML documents in XML DB is added, that is, 
\texttt{FOR} clause for this purpose is added at the front of this instance.
Second, document loading function $\mathit{doc}(\cdot)$ is replaced to the 
variable in the added \texttt{FOR} clause.
Third, boolean condition in \texttt{RETURN} is moved to \texttt{WHERE} clause.
If \texttt{WHERE} clause does not exist in this instance, it is appended,
otherwise this condition is added to \texttt{WHERE} clause with conjunction keyword (i.e., \texttt{and}).
At last, expression in \texttt{RETURN} clause is replaced
to $\mathit{base\mbox{-}uri}(\cdot)$\footnote{
\url{https://www.w3.org/TR/xpath-functions-31/#func-base-uri}}
for returning identifiers of XML documents.
For instance, XQuery in Listing~\ref{lst:xfilter} is rewritten as 
Listing~\ref{lst:rewriting} (red-colored parts are changed by the rules above).
The first line is added by the first rule where `safety\_info' is a name of collection in XML DB,
variable $\$$doc in the second line is replaced from $doc(?doc)$ by the second rule.
\texttt{WHERE} clause is added with the condition in \texttt{RETURN} clause by the third rule, and 
the expression in \texttt{RETURN} clause is replaced with $\mathit{base\mbox{-}uri}(\$doc)$
which returns identifiers of XML documents bound to $\$$doc variable by the last rule.

%\begin{figure}[b]
\begin{lstlisting}[
	mathescape=true,
	caption=Example of XQuery Rewriting in Parallel plan (red: changes),
	label=lst:rewriting,
	numbers=left,
	numbersep=-8pt
]
    (*@\texttt{\color{red} \textbf{FOR} $\$$doc in collection('safety\_info')}@*)
    LET $\$$x := (*@\texttt{\color{red} $\$$doc}@*)//mail[leaveDate > xs:date('2020-03-01')]
    (*@\texttt{\color{red} \textbf{WHERE}}@*) contains($\$$x, 'coronavirus')
    RETURN (*@\texttt{\color{red} base-uri($\$$doc)} @*)
\end{lstlisting}
%\end{figure}

To generate query result of SPARQL with XQuery-based filtering, respective results of
SPARQL and XQuery are joined in an equi-join manner, 
since these results contain XML document identifiers in common.
Variable corresponding with these identifiers is used as join key to find bindings of
SPARQL variables that satisfy XQuery conditions.
Since the join process is done on memory, hash join is a reasonable option of join algorithm.
If the results do not fit in memory, sort-merge join or other algorithms can be considered\footnote{In
this paper, enough amount of memory is assumed, thus to chose
the best join algorithm when results do not fit in memory is out of scope of this paper. }.

\subsection{SPARQL First}

Basic idea of this query plan is that, when the number of binding in SPARQL instance is small,
only the bindings are necessary to be examined in XQuery evaluation, therefore,
the cost for XQuery processing can be reduced.
To this end, similar to the parallel plan, SPARQL instance is rewritten to add 
the SPARQL variable in XQuery instance into its \texttt{SELECT} clause.

To push the bindings down to XQuery instance, this instance is rewritten by a similar rule
to one in the parallel plan. 
In this rule, first, document enumeration on the basis of the bindings is added, that is,
\texttt{FOR} clause including a list of XML documents in the bindings is added at the front of this instance.
Second, document loading function $\mathit{doc}(\cdot)$ is replaced to the variable 
in the added \texttt{FOR} clause in the same way as the parallel query plan.
At last, the expression in \texttt{RETURN} clause is modified to represent tuples of 
XML document identifier and boolean value of whether the document satisfies condition.
Listing~\ref{lst:rewriting_sparql} showcases an example of rewritten XQuery instance
of the query in Listing~\ref{lst:xfilter}.
The first line is added by the first rule and the second line is modified in accordance with the second rule.
The \texttt{RETURN} clause is changed as the third to forth lines for pairing
document ID and boolean value.

%\begin{figure}[t]
\begin{lstlisting}[
	mathescape=true,
	caption=Example of XQuery Rewriting in SPARQL First plan (red: changes),
	label=lst:rewriting_sparql,
	numbers=left,
	numbersep=-8pt
]
    (*@\texttt{\color{red} \textbf{FOR} $\$$doc in ('0001.xml', '0002.xml', ...)}@*)
    LET $\$$x := (*@\texttt{\color{red} $\$$doc}@*)//mail[leaveDate > xs:date('2020-03-01')]
    RETURN (*@\texttt{\color{red} <tuple><doc>\{$\$$doc\}</doc>}@*)
                  (*@\texttt{\color{red} <bool>\{}contains($\$$x, 'coronavirus'){\color{red} \}</bool></tuple>}@*) 
\end{lstlisting}
%\end{figure}

To generate query result of SPARQL with XQuery-based filtering, bindings of SPARQL variables is filtered 
by the results of XQuery.
This process can also be a join processing, and again this join processing is equi-join,
therefore, hash join is performed.
In contrast to the parallel plan, the size of XQuery results is much smaller, thus,
join cost is also much smaller than that in the parallel plan.

\subsection{XQuery First}

Basic idea of this query plan is similar to that of the SPARQL first plan,
that is, when the number of bindings in XQuery instance is small,
only the bindings are necessary to be examined in SPARQL evaluation, therefore,
the cost for SPARQL processing can be reduced.
This is effective when SPARQL instance includes \texttt{SERVICE} clause for
accessing remote SPARQL endpoints, which requires higher query processing cost.
Since XQuery is used to enumerate XML documents which satisfy the conditions in XQuery,
rewriting XQuery in this plan is exactly same way as that in the parallel plan.

To push results of XQuery instance down to SPARQL instance, the SPARQL instance
is modified as a sequence of \texttt{UNION} operators over graph patterns corresponding
with the XQuery results.
Each graph pattern is a replica of \texttt{WHERE} clause of the original SPARQL instance,
except replacing the variable appearing also in XQuery instance to one of these results.
Listing~\ref{lst:rewriting_xquery} displays a rewriting example of SPARQL with
XQuery-based filtering in Listing~\ref{lst:xfilter}.
\texttt{WHERE} clause in the original SPARQL instance is replicated for graph patterns,
and these graph patterns are concatenated \texttt{UNION} operators.
Variable \texttt{?doc} is replace with identifier for XML document.
Result set of the rewritten SPARQL instance is equivalent to that of the original 
query of SPARQL with XQuery-based filtering, therefore, no post-processing is required in this plan. 

\begin{figure}[t]
\begin{lstlisting}[
	mathescape=true,
	caption=Example of XQuery Rewriting in XQuery First plan (red: changes),
	label=lst:rewriting_xquery,
	numbers=left,
	numbersep=-8pt
]
    SELECT ?s
    WHERE {
      (*@\texttt{{\color{red} \{} ?s rdf:type ex:Country ; ex:safetyInfo  {\color{red} '0001.xml'} ; owl:sameAs ?x .}@*)
        SERVICE <http://dbpedia.org/sparql> {
           SELECT ?x
           WHERE { ?x dbo:populationTotal ?pop .
                   FILTER ( ?pop > 10,000,000 ) . 
      } } (*@\texttt{\color{red} \}}@*)
      (*@\texttt{\color{red}UNION}@*)
      (*@\texttt{{\color{red} \{} ?s rdf:type ex:Country ; ex:safetyInfo {\color{red} '0002.xml'} ; owl:sameAs ?x .}@*)
        SERVICE <http://dbpedia.org/sparql> {
           SELECT ?x
           WHERE { ?x dbo:populationTotal ?pop .
                   FILTER ( ?pop > 10,000,000 ) . 
      } } (*@\texttt{\color{red} \}}@*)
      (*@\texttt{\color{red}UNION}@*)
      ...
    }
\end{lstlisting}
\end{figure}

\subsection{Optimization}
\label{sec:prop_opt}

Basic idea of query optimization is to chose a query plan with the least cost for execution.
To this end, execution costs of query plans are necessary to be modeled.
Execution cost of the parallel plan depends on the worst performance of SPARQL endpoint or XML DB
and on the join cost of post-processing.
Execution cost of the SPARQL first plan is summation of SPARQL query processing cost,
reduced XQuery cost by SPARQL result pushdown and the join cost.
Similarly, execution cost of the XQuery first plan is summation of XQuery processing cost,
reduced SPARQL cost by XQuery result pushdown.

Let $C^{(p)}$, $C^{(s)}$ and $C^{(x)}$ denote costs of the parallel, SPARQL first and
XQuery first plans, $C_{\mathit{SPARQL}}$ and $C_{\mathit{XQuery}}$ respectively denote
the solo processing costs of SPARQL and XQuery, and $C_{\mathit{Join}}^{(p)}$ and 
$C_{\mathit{Join}}^{(s)}$ respectively denote the join costs in the parallel plan
and in the SPARQL first plan. 
On the basis of these notations, the costs of the three plans are
formalized as Equations~\ref{eq:cost_parallel}, \ref{eq:cost_sparql} and \ref{eq:cost_xquery}.
The amount of cost reduction on the subsequent query processing in the SPARQL first and 
XQuery first plans are represented by selectivity ratios,
$\rho_{\mathit{SPARQL}}$ and $\rho_{\mathit{XQuery}}$, of preceding SPARQL and XQuery instances.
\begin{align}
 C^{(\mathit{p})} &= \max(C_{\mathit{SPARQL}}, C_{\mathit{XQuery}}) + C_{\mathit{Join}}^{(p)} 
 	\label{eq:cost_parallel} \\
 C^{(\mathit{s})} &= C_{\mathit{SPARQL}} + \rho_{\mathit{SPARQL}} \cdot C_{\mathit{XQuery}}  + C_{\mathit{Join}}^{(s)}
 	\label{eq:cost_sparql} \\
 C^{(\mathit{x})} &= C_{\mathit{XQuery}} + \rho_{\mathit{XQuery}} \cdot C_{\mathit{SPARQL}}
 	\label{eq:cost_xquery}
\end{align}

To explore the optimal plan, in this paper, these individual costs and selectivity 
ratios in these equations are estimated using statics from query execution history
and conventional selectivity estimation techniques for SPARQL (e.g., \cite{rdf3x,selectivity_sparql_0})
and XQuery (e.g., \cite{selectivity_xquery_0,selectivity_xquery_1}).
The main objective of this paper in this optimization is to model the costs of 
query plans, and estimating these costs and selectivity ratios are out of scope.
Therefore, in this paper, the costs and selectivity ratios are assumed to be ideally estimated.

%% file: src/exp.tex
\section{Experimental Evaluation}
\label{sec:exp}

In this paper, experimental evaluation to show efficiency of 
the proposed SPARQL with XQuery-based filtering technique and query optimization is conducted.
In this experiment, three real-world scenarios are prepared.
These scenarios have different settings of databases in terms of network latency and data size.
Datasets and queries are explained in the subsequent section.
For XML data, local XML database is prepared and XML data are stored in it.
For RDF data, local and external SPARQL endpoints are used.
For the local databases, eXist-db (version 5.2.0)\footnote{\url{http://exist-db.org/}} and 
Apache Jena Fuseki (version 3.14.0)\footnote{\url{https://jena.apache.org/documentation/fuseki2/index.html}} 
are used for XML DB and SPARQL endpoint, respectively.
In this experiment, query execution time is used for efficiency measurements
of query processing.

\subsection{Scenarios}

\begin{table}[t]
	\centering
	\begin{minipage}{.49\textwidth}
		\centering
		\caption{Expected DB Performance}
		\begin{tabular}{c!{\vrule width .15em}c|c}
			\specialrule{.15em}{.15em}{.15em}
			\multirow{2}{*}{Scenario} & \multicolumn{2}{c}{Database} \\
			\cline{2-3}
			 & SPARQL EP & XML DB \\
			\specialrule{.15em}{.15em}{.15em}
			C.S. & low & high \\
			L.S. & comparable & comparable \\
			D.S. & high & low \\
			\specialrule{.15em}{.15em}{.15em}
		\end{tabular}
		\label{tab:db_performance}
	\end{minipage}
	\hfill
	\begin{minipage}{.49\textwidth}
		\centering
		\caption{Data Size for Scenarios}
		\begin{tabular}{c!{\vrule width .15em}r|r}
			\specialrule{.15em}{.15em}{.15em}
			\multirow{2}{*}{Scenario} &\multicolumn{2}{c}{Data Type} \\
			\cline{2-3}
			& XML (\#doc) & \multicolumn{1}{c}{LOD (\#ent)} \\
			\specialrule{.15em}{.15em}{.15em}
				C.S. & 207 & 207 + DBpedia \\
				L.S. & 8,435 & 106,341 \\ 
				D.S. & 107,193 & 317,552 \\
			\specialrule{.15em}{.15em}{.15em}
		\end{tabular}
		\label{tab:dataset}
	\end{minipage}
\end{table}

In this experiment, three scenarios are prepared in order to observe query performances 
over different performances of underlying databases as summarized in Table~\ref{tab:db_performance}.
The first scenario (C.S. scenario) is to search countries with governmental safety information messages. 
This scenario is the case that cost of querying to SPARQL endpoint is high and 
that of querying to XML DB is low. 
This case is realized by using external SPARQL endpoint via \texttt{SERVICE} clause
and small number of XML documents.
The second scenario (L.S. scenario) is to search acts with their body texts. 
This scenario is the case that costs of querying SPARQL and XQuery are comparable.  
This case is realized by local SPARQL endpoint and moderate number of XML documents.
The third scenario (D.S. scenario) is to search discussions with minute books about law enactment.
This scenario is the case that cost of querying SPARQL is low and that of querying 
XQuery is high.
This case is realized by local SPARQL endpoint and large number of XML documents.
The following summarize datasets and queries in these scenarios,
and Table~\ref{tab:dataset} showcases dataset sizes. 
Queries are generated by using templates to control the selectivities 
of SPARQL and XQuery instances\footnote{Due to the space limit,
the template of only the C.S. scenario is displayed.}.

\textbf{Scenario (C.S.) -- Country Search with Safety Information}:
XML data in this scenario is Overseas Travel Safety Information Open
Data\footnote{\url{https://www.ezairyu.mofa.go.jp/html/opendata/index.html} (in Japanese)}
by Ministry of Foreign Affairs (MOFA) of Japan.
Each XML document corresponds to a country and consists of a bunch of past messages about 
safety information of the corresponding country with dates.
For RDF data, DBpedia\footnote{\url{http://dbpedia.org/}} is used for referring to country information 
(e.g., population), and to connect these XML data with country entities in DBpedia, mapping
relations between safety information XML data and corresponding countries in DBpedia 
are prepared as a local SPARQL endpoint. 
On the basis of these data, queries are designed to search countries having more population 
than specified and there are warning messages about a particular keyword before specified date. 
Query template is shown in Listing~\ref{lst:template_cs}.
In the following query templates, `\_pop\_' and `\_date\_' are placeholders 
for conditions in order to control selectivity.

\begin{figure}[t]
\begin{lstlisting}[
	basicstyle=\ttfamily,
	mathescape=true,
	caption=C.S. Query Template -- Countries with more than `\_pop\_' people and with at least one announcement about coronavirus before `\_date\_',
	label=lst:template_cs
]
  SELECT ?entity ?safety
  WHERE { ?entity ex:safetyInfo ?safety .
    SERVICE <http://dbpedia.org/sparql> {
       ?entity dbo:populationTotal ?pop .
       FILTER (?pop > _pop_ ) .
    }
    XQueryFILTER (
       LET $\$$d := doc(?safety)//mail[leaveDate < xs:date('_date_')]
       RETURN contains($\$$d, 'coronavirus')
    ) .
  }
\end{lstlisting}
\end{figure}

\textbf{Scenario (L.S.) -- Law Search with Body Text}:
XML data in this scenario is law body texts from e-Gov\footnote{\url{https://www.e-gov.go.jp/}}, 
a portal site for administrative information, by Ministry of Internal Affairs and Communications of Japan.
Each XML document corresponds with a law.
For RDF, Japanese law history LOD~\cite{ckg2019} that contains 106,341 laws is used.
In this LOD, some law entities have a link to XML documents of their body text. 
Using these data, queries are designed to analyze sizes of laws in different era.
%Query template is shown in Listing~\ref{lst:template_ls}.

%\begin{figure}[t]
%\begin{lstlisting}[
%	basicstyle=\ttfamily,
%	mathescape=true,
%	caption=L.S. Query Template -- Acts with more than yyy articles after xxx,
%	label=lst:template_ls
%]
%  SELECT ?lawid
%  WHERE { ?lawid rdf:type lh_lod:Act ; ndl_prop:eGovID ?egov ;
%           lh_lod:promulgateDate ?d .
%    FILTER (?d > xs:date('xxx')) .
%    XQueryFILTER (
%       LET $\$$d := doc(?egov)
%       LET $\$$c := count($\$$d//MainProvision//Article)
%       RETURN $\$$c > yyy
%    ) .
%  }
%\end{lstlisting}
%\end{figure}

\textbf{Scenario (D.S.) -- Discussion Search with Minute Book}:
XML data in this scenario is minute books about law enactments from a minute book search
service\footnote{\url{https://kokkai.ndl.go.jp/} (in Japanese)} provided by National Diet Library (NDL) of Japan. 
The XML data includes 107,193 minute books.
For LOD, Japanese law history LOD~\cite{ckg2019} is again used in this scenario, 
and it contains 317,552 links to minute books.
On these data, queries are designed to search laws with minute books that a particular person speaks 
and their conference is held before specified date. 
%Query template is shown in Listing~\ref{lst:template_ds}.

%\begin{figure}[t]
%\begin{lstlisting}[
%	basicstyle=\ttfamily,
%	mathescape=true,
%	caption=D.S. Query Template -- Acts that yyy spoke in discussions before xxx,
%	label=lst:template_ds
%]
%  SELECT distinct ?lawid ?disc
%  WHERE {
%    ?lawid rdf:type lawhist:Act ; ndl_diss:hasHistory ?hist .
%    ?hist  ndl_diss:hasDiscussion ?disc .
%    ?disc  ndl_diss:conferenceDate ?date ; ndl_diss:minuteID ?minute .
%    FILTER (?date < xs:date('xxx')) .
%    XQueryFILTER (
%       LET $\$$d := doc(?minute)
%       LET $\$$c := $\$$d/meetingRecord/speechRecord/speaker
%       RETURN contains($\$$c, 'yyy')
%    ) .
%  }
%\end{lstlisting}
%\end{figure}

\subsection{Results}

\begin{figure}[t]
	\centering
	\subfloat[C.S. (Country Search)]{
		\centering
		\includegraphics[width=.32\textwidth]{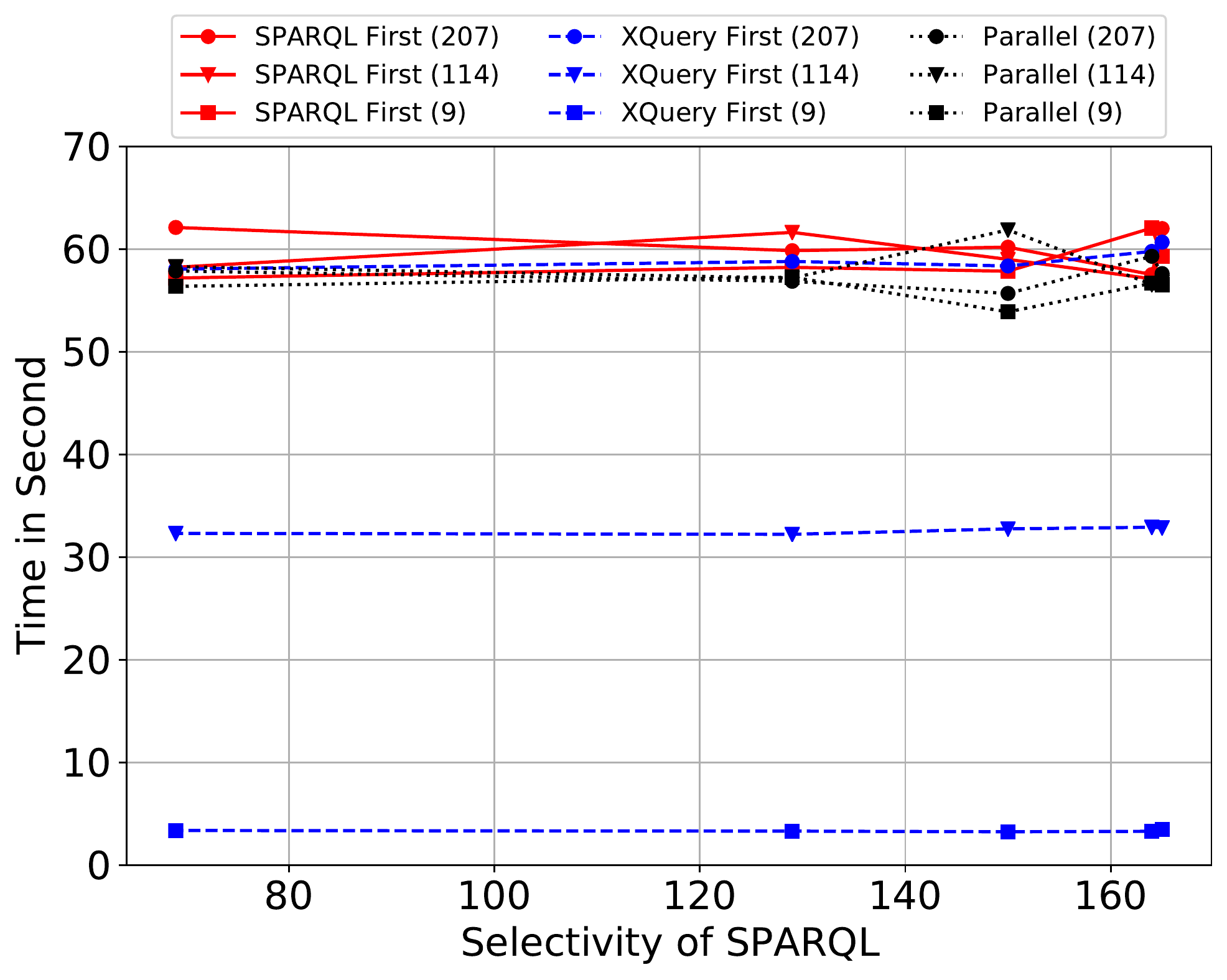}	
		\label{fig:safety}
	}
	\subfloat[L.S. (Law Search)]{
		\centering
		\includegraphics[width=.32\textwidth]{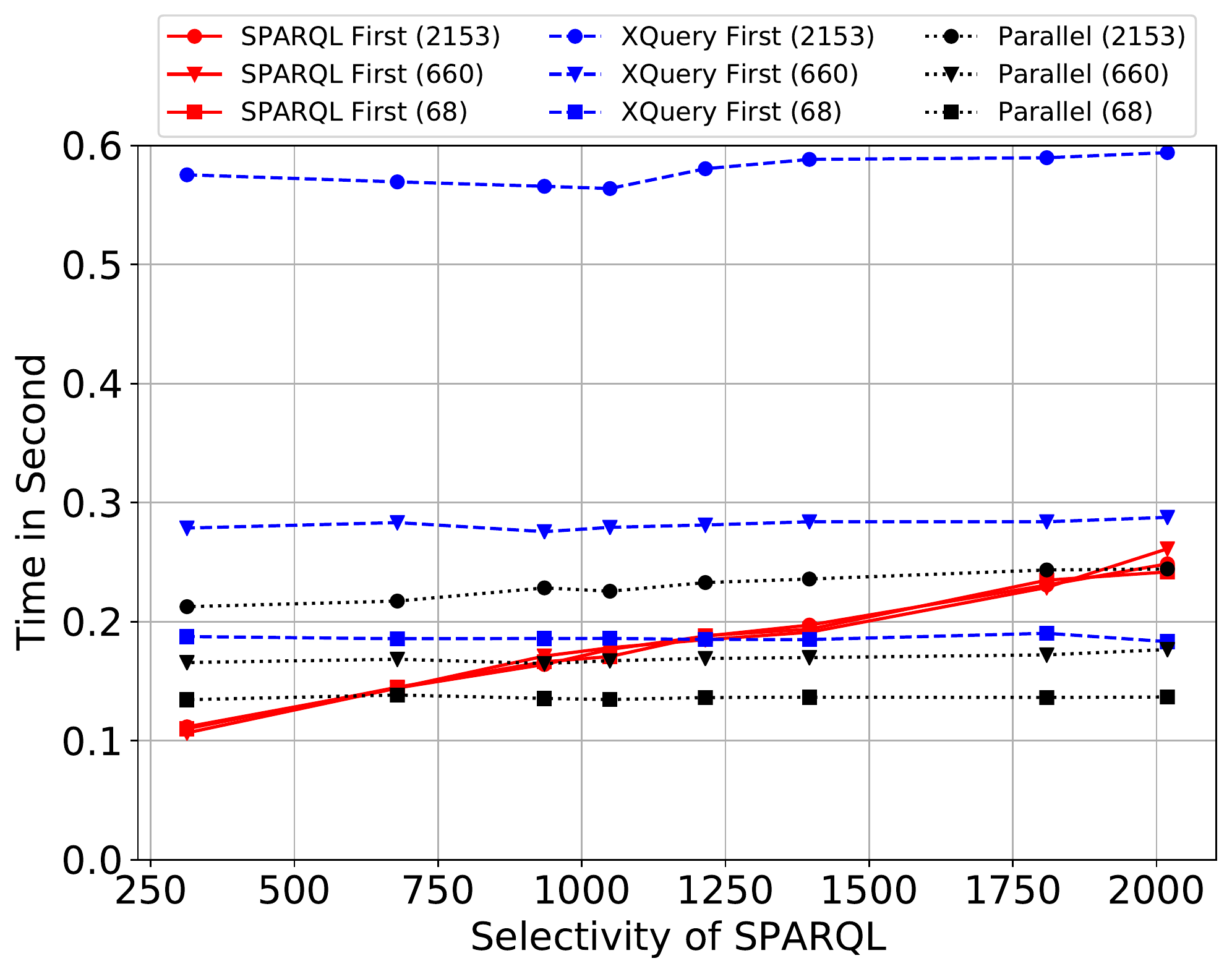}	
		\label{fig:law_doc}
	}
	\subfloat[D.S. (Discussion Search)]{
		\centering
		\includegraphics[width=.32\textwidth]{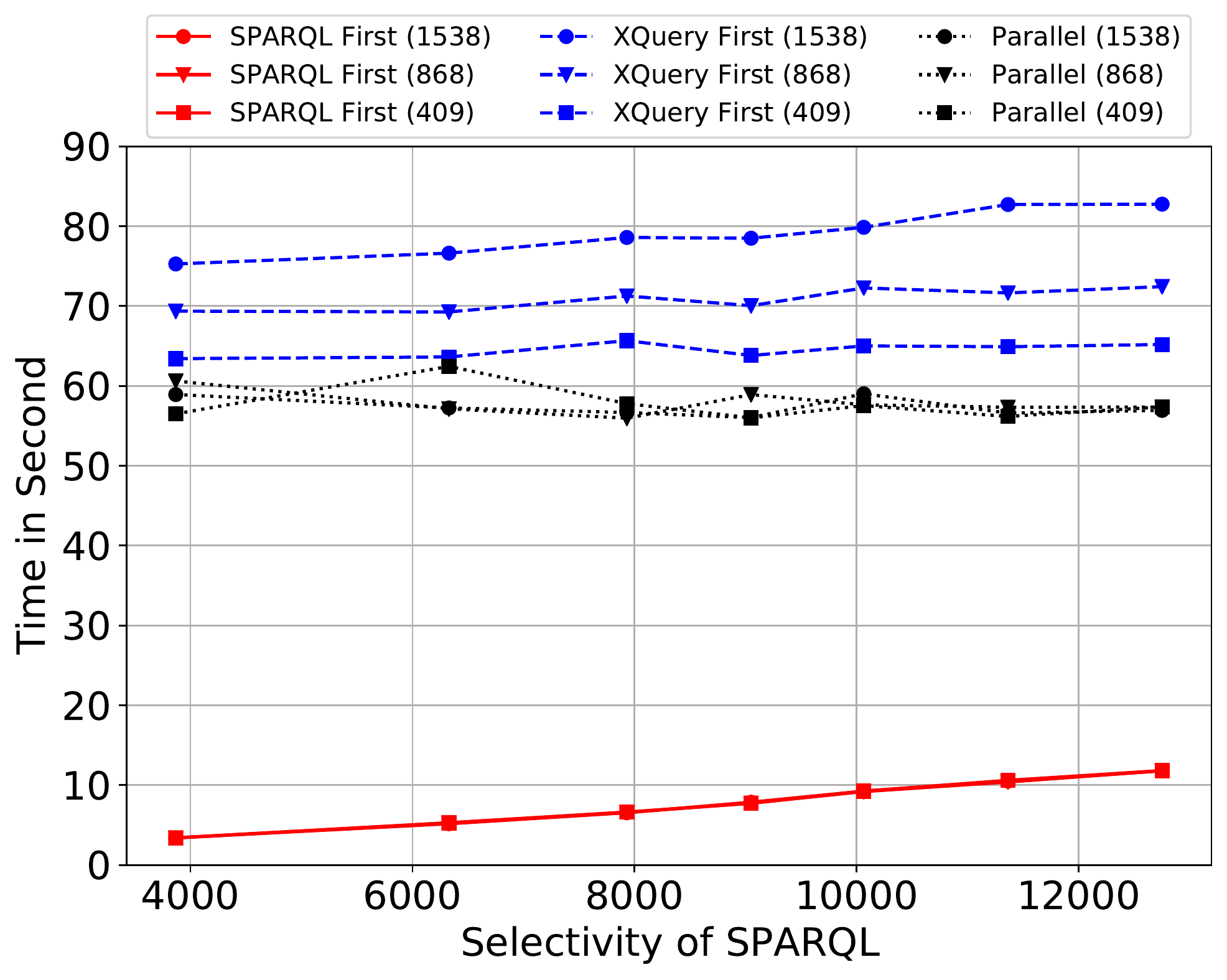}
		
		\label{fig:minute}
	}
	\caption{Execution Time in Scenarios (Number in Legend: XQuery Selectivity)}
	\label{fig:time_overall}
\end{figure}

Figure~\ref{fig:time_overall} shows selected results for each scenario 
in terms of selectivities of SPARQL and XQuery.
The lines in figures in Figure~\ref{fig:time_overall} represent execution times
over SPARQL selectivity (i.e., the number of results of solo SPARQL instance).
There are three lines for each query plan (query plans are colored in different styles,
that is, red solid for SPARQL first, blue dashed for XQuery first and black dotted for 
Parallel), which correspond with high, middle and low XQuery selectivities, respectively.
They are represented by circle, triangle and square markers, respectively.
The common observation from Figure~\ref{fig:time_overall} is as follows.
First, the SPARQL first plan (resp. XQuery first plan) is linearly performed
in terms of selectivity of SPARQL (resp. XQuery) query and it is scarcely affected
by selectivity of XQuery (resp. SPARQL) query.
Second, the parallel plan is nearly constant in terms of both selectivities of 
SPARQL and XQuery instances when their execution performances have a large gap like
C.S. and D.S. scenarios.
If these performances are comparable (as the L.S. scenario), this plan depends 
on both selectivities of SPARQL and XQuery as shown in Figure~\subref*{fig:law_doc}.

As summarized in Table~\ref{tab:db_performance}, query performances on these scenarios
are expected varying, and figure~\ref{fig:time_overall} demonstrates different query
execution behaviors as expected in the cost models in Section~\ref{sec:prop_opt}.
When SPARQL endpoint has low performance (i.e., C.S. scenario in Figure~\subref*{fig:safety}),
the XQuery first plan is clearly superior to other plans, especially when 
XQuery selectivity is low. 
On the contrary,
when XML DB has low performance (i.e., D.S. scenario in Figure~\subref*{fig:minute}),
the SPARQL first plan is always superior to the other plans.
For the comparable scenario (i.e., L.S. scenario in Figure~\subref*{fig:law_doc}),
the parallel and the SPARQL first plans are comparable.

\begin{figure}[t]
	\centering
	\subfloat[Parallel]{
		\centering
		\includegraphics[width=.32\textwidth]{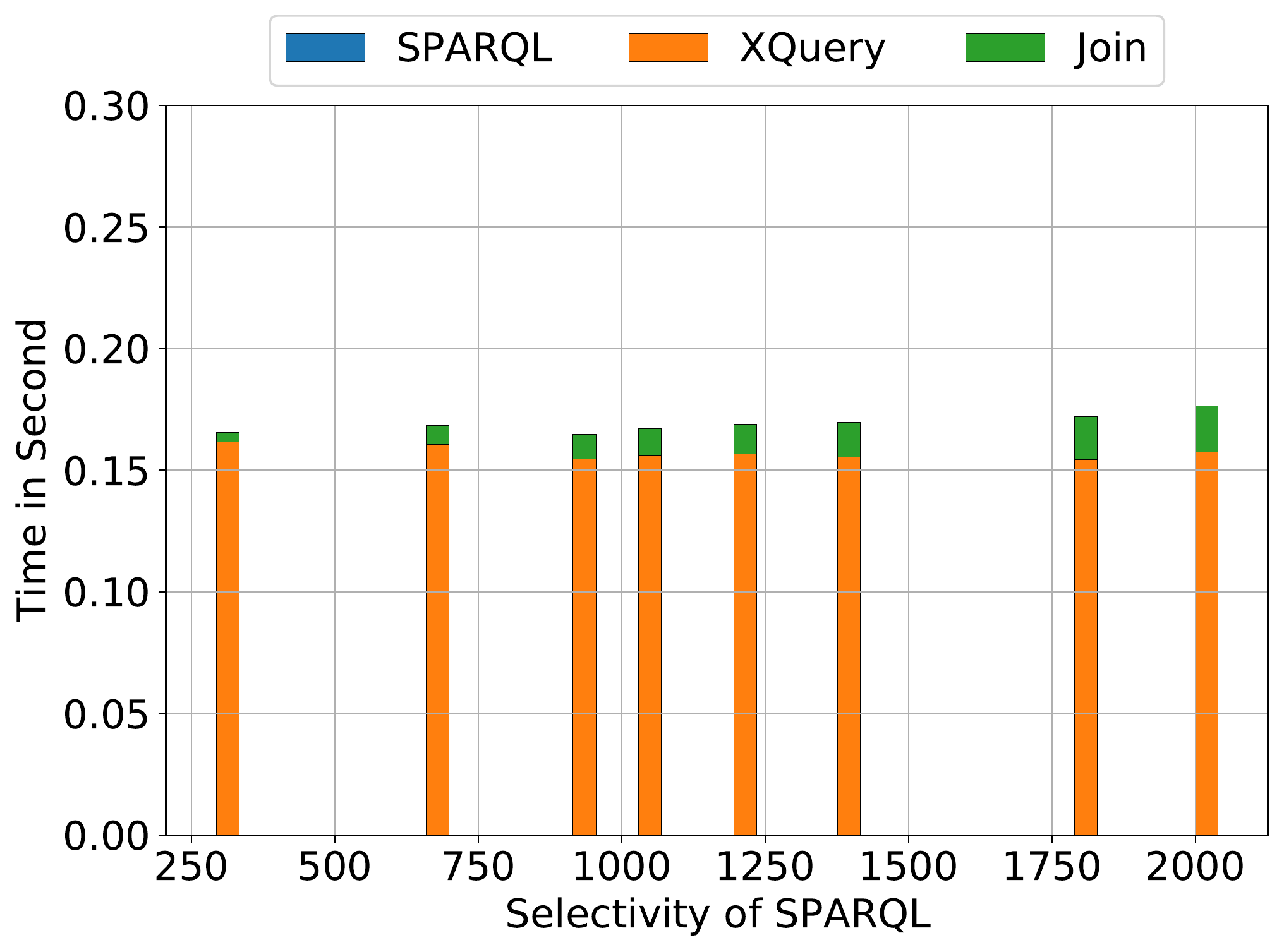}	
		\label{fig:parallel}
	}
	\subfloat[SPARQL First]{
		\centering
		\includegraphics[width=.32\textwidth]{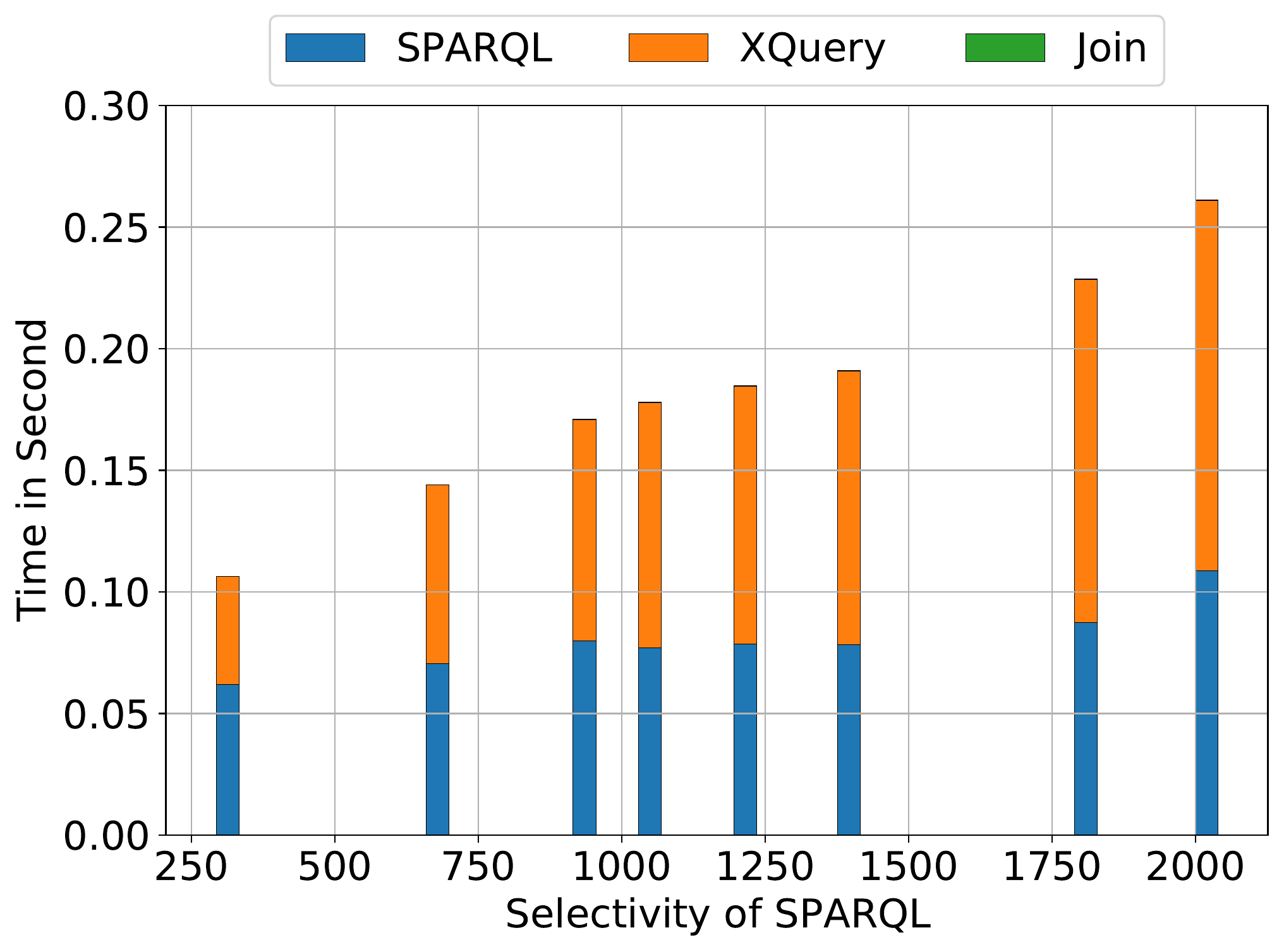}	
		\label{fig:sparql}
	}
	\subfloat[XQuer First]{
		\centering
		\includegraphics[width=.32\textwidth]{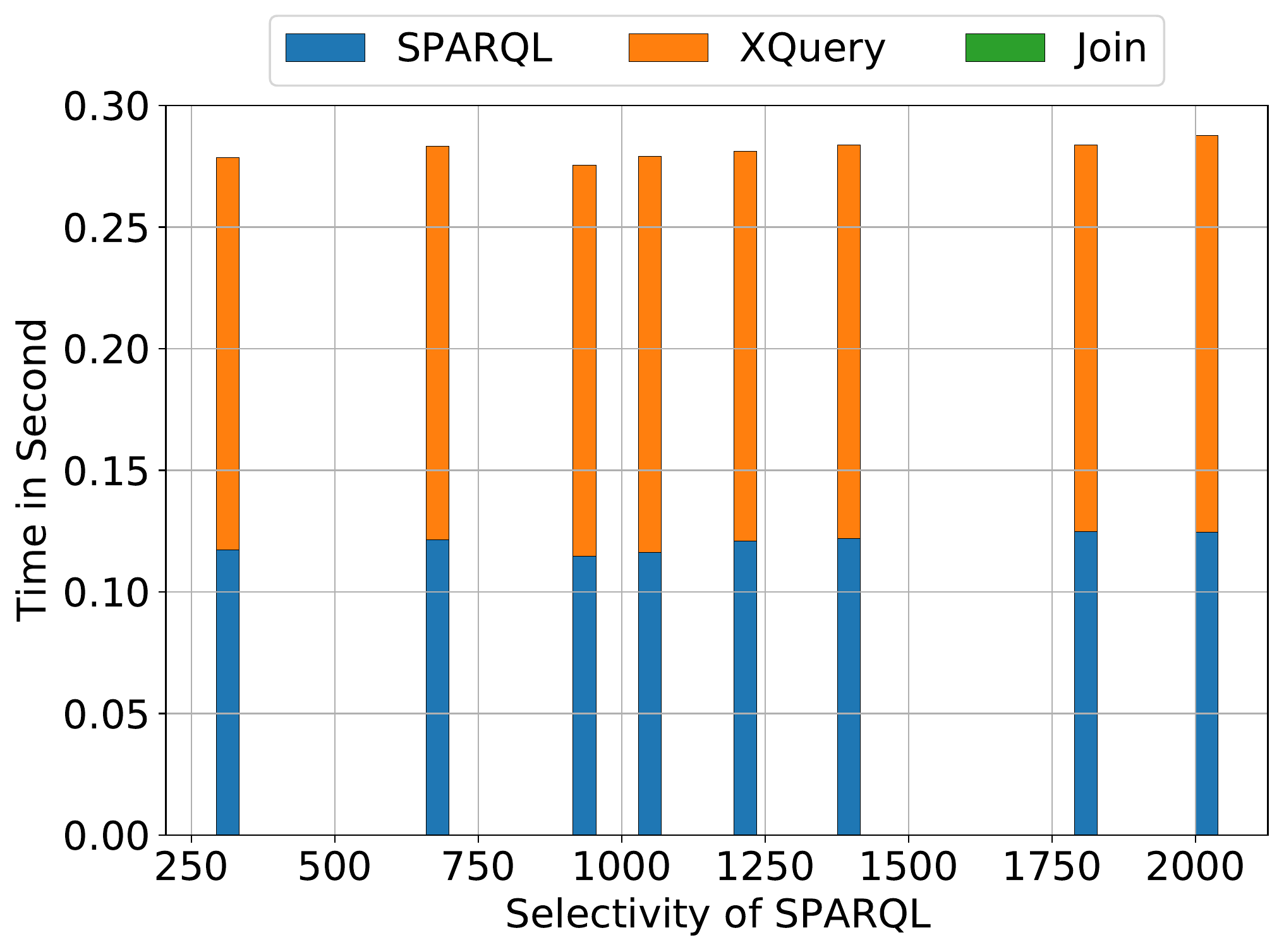}	
		\label{fig:xquery}
	}
	\caption{Detail of Execution Time in L.S. Scenario with XQuery Selectivity of 660}
	\label{fig:time_detail}
\end{figure}

Figure~\ref{fig:time_detail} shows a breakdown of execution time for each module in 
query plans (Figure~\ref{fig:manager}).
This figure is a stacked bar graph of execution times of SPARQL, XQuery and Join,
which are illustrated with blue, orange and green colors.
Theses figures are about a particular condition in the L.S. scenario where 
XQuery selectivity is 660 (triangular marker in Figure~\subref*{fig:law_doc}).
One basic fact in Figure~\ref{fig:time_detail} is that join times in both the parallel 
and SPARQL first plans have few effect on their overall performances.

In Figure~\subref*{fig:law_doc}, the optimal plan for XQuery selectivity of 660 is switched
from the SPARQL first plan to the parallel plan around SPARQL selectivity of 1,000.
Figure~\ref{fig:time_detail} indicate a reason for this switching.
Basically, in this SPARQL with XQuery-based filtering query, SPARQL instance is executable
relatively faster than XQuery (this can be seen from SPARQL execution time in 
Figure~\subref*{fig:sparql} and XQuery execution time in Figure~\subref*{fig:xquery}).
Therefore, SPARQL first plan is better plan as far as the number of XML documents
being queried afterward is small.
In this scenario, the number of queried XML documents more than 1,000 is a turning point 
that the parallel plan overcomes the SPARQL first plan.

\subsection{Lessons Learned}

This experiment indicates that \texttt{XQueryFILTER} is useful when XML data 
is associated with entities in LOD.
The users who are knowledgable to both SPARQL and XQuery can easily use 
SPARQL with XQuery-based filtering.
Furthermore, the three query plans modeling in this paper reasonably reflect 
the pros and cons of underlying databases, namely, SPARQL endpoint and XML DB.
On the basis of settings of these databases (e.g., network latency and data size),
these plans have best-performing conditions. 
When SPARQL endpoint (resp. XML DB) performs far better than XML DB (resp. SPARQL endpoint),
the SPARQL first plan (resp. XQuery first plan) is preferred because results of
SPARQL (resp. XQuery) can reduce the processing cost of the subsequent XQuery
(resp. SPARQL) query processing.
For the intermediate case when performances of SPARQL endpoint and XML DB are 
comparable, the parallel plan can perform better on average, 
however, when selectivities of SPARQL and XQuery instances are both smaller,
the SPARQL first plan is preferred.
The proposed optimization technique captures these characteristics of plans by
the cost equations in Equation~\ref{eq:cost_parallel}, \ref{eq:cost_sparql} and \ref{eq:cost_xquery}
and can successfully discover the best plan if statistics of database performances and 
selectivity estimations of SPARQL and XQuery are accurate.

%% file: src/con.tex
\section{Conclusion and Future Direction}
\label{sec:con}

\textbf{Conclusion}:
In this paper, an extension of SPARQL which incorporates XQuery 
as a filtering mechanism, \texttt{XQueryFILTER}, to handle XML
documents connected with entities in Linked Open Data is proposed.
In contrast to data-centric XML data, document-centric XML data needs effort
to construct LOD, therefore, it is preferable to handle XML data intact.
To realize \texttt{XQueryFILTER}, in this paper, three query plans 
are proposed and a database theory-based query optimization is proposed
on the basis of selectivity estimations on SPARQL and XQuery.
The cost models for these query plans reflect their behaviors based on 
the query execution costs of SPARQL endpoint and XML DB as well as 
selectivity ratios of SPARQL and XQuery instance.
In the experimental evaluation on real-world scenarios, 
the optimization strategy successfully chose the optimal strategy.

\noindent
\textbf{Future Directions}:
There are three promising future directions.
First, the proposed filter can be extended to deal with other types of 
structured documents like JSON by applying similar idea.
Second, the proposed filter can be relaxed for general-purpose 
XQuery processing such as simultaneously querying LOD and XML documents.
Last, to improve usability, user-friendly queries are preferable
such as keyword search and faceted search, which are extensively 
studied solely on RDF data and XML data (keyword search on RDF
data~\cite{keyword_rdf} and XML data~\cite{keyword_xml}, and 
%faceted search on RDF data~\cite{faceted_rdf1,faceted_rdf2} 
faceted search on RDF data~\cite{faceted_rdf1}
and XML data~\cite{iiwas2011}).